\documentclass[twocolumn]{aastex7}

\usepackage{subfigure}
\usepackage{color}
\usepackage{comment}
\usepackage{url}
\usepackage{graphicx}
\usepackage{natbib}

\newcommand{\etaTel}{$\eta$~Tel}

\newcommand{\OI}{\ion{O}{1}}
\newcommand{\CI}{\ion{C}{1}}
\newcommand{\CII}{\ion{C}{2}}
\newcommand{\NI}{\ion{N}{1}}
\newcommand{\SII}{\ion{S}{2}}
\newcommand{\SI}{\ion{S}{1}}
\newcommand{\FeII}{\ion{Fe}{2}}
\newcommand{\AlII}{\ion{Al}{2}}
\newcommand{\SiII}{\ion{Si}{2}}
\newcommand{\MnII}{\ion{Mn}{2}}
\newcommand{\MgII}{\ion{Mg}{2}}
\newcommand{\kms}{km s$^{-1}$}

\defcitealias{Youngblood2021}{Y21}

\begin{document}

\title{Revisiting the ultraviolet spectroscopy of the \etaTel\ edge-on debris disk}

\author[0000-0002-1176-3391]{Allison Youngblood}
\affiliation{Exoplanets and Stellar Astrophysics Laboratory, NASA Goddard Space Flight Center, Greenbelt, MD 20771, USA}
\email{allison.a.youngblood@nasa.gov}

\author[0000-0002-7201-7536]{Alexis Brandeker}
\affiliation{AlbaNova University Centre, Stockholm University, Department of Astronomy, Stockholm, Sweden}
\email{alexis@astro.su.se}

\author[0000-0003-2953-755X]{Sebasti\'an P\'erez}
\affiliation{Departamento de F\'isica, Universidad de Santiago de Chile. Avenida Ecuador 3493, Estaci\'on Central, Santiago, Chile}
\affiliation{Center for Interdisciplinary Research in Astrophysics and Space Exploration (CIRAS), Universidad de Santiago de Chile, Chile}
\affiliation{Millennium Nucleus on Young Exoplanets and their Moons (YEMS), Chile}
\email{sebastian.astrophysics@gmail.com}

\author[0000-0002-2989-3725]{Aki Roberge}
\affiliation{Exoplanets and Stellar Astrophysics Lab, NASA Goddard Space Flight Center, Greenbelt, MD 20771, USA}
\email{aki.roberge-1@nasa.gov}

\author[0000-0001-6654-7859]{Alycia Weinberger} 
\affiliation{Department of Terrestrial Magnetism, Carnegie Institution for Science, 5241 Broad Branch Road NW, Washington, DC 20015, USA}
\email{aweinberger@carnegiescience.edu}

\author[0000-0001-7891-8143]{Meredith A. MacGregor} 
\affiliation{Center for Astrophysics and Space Astronomy, University of Colorado, 389 UCB Boulder, CO 80309, USA}
\email{mmacgregor@jhu.edu}

\author{Barry Welsh}
\affiliation{Eureka Scientific, 2452 Delmer, Suite 100, Oakland, CA 96002, USA}
\email{barryywelsh@yahoo.com}

\begin{abstract}

We revisit the ultraviolet absorption spectroscopy of the edge-on debris disk surrounding the A0V star $\eta$~Telescopii. Previous work found absorption components at four velocities ($\sim$ -23, -18, -10, -1 \kms), with the most blueshifted component { (-23 \kms)} interpreted as a likely disk wind. However, optical spectroscopy of \etaTel\ and other nearby stars in projection demonstrate that the -23 \kms\ component is likely interstellar in origin. We find that there are three interstellar components toward this sight line (-23, -18, -10 \kms), but that the fourth component near -1 \kms, which was only detected in \OI, is inconsistent with an interstellar origin and could be circumstellar. We place a 3-$\sigma$ upper limit on the C/O ratio of the -1 \kms\ gas (log C/O $<$ -2.1), finding that it is consistent with Earth and solar system comet abundances. { However, the abundance is inconsistent with the carbon-rich disks of $\beta$~Pic (A5V) and 49 Cet (A1V), probably because \etaTel\ (A0V) is a warmer star imposing greater levels of radiation pressure on carbon atoms in the disk. } A low C/O ratio is also inconsistent with Herschel's [\CII] detection toward \etaTel\ and may indicate that carbon gas is misaligned from the line of sight or variable in time. 

\end{abstract}


\section{Introduction} \label{sec:Introduction}


{ Compared to protoplanetary disks, debris disks are generally gas poor in terms of total gas mass, with any primordial gas remaining from the protoplanetary stage cleared out by the star or depleted during planet formation. Gas in these disks is continually sourced through collisions between planetesimals  \citep{Beust:1990,Czechowski:2007,Grigorieva:2007,Zuckerman:2012}, although a primordial origin has been proposed for some CO-rich debris disks \citep{Kospal2013,Moor2017}. Assuming a secondary origin, characterizing the gas of debris disks reveals the chemical composition of planetesimals. }

\etaTel\ A is a young A0V star at 48.5 pc \citep{GaiaDR3} in the $\sim$23 Myr old $\beta$~Pic moving group \citep{Zuckerman2001,Mamajek2014}. It has a brown dwarf companion, \etaTel\ B,  approximately 208 au away and a third comoving star, HD 181327 (F6V), which is approximately 20,000 au away \citep{Nogueira2024} and has a face-on debris disk ($i\approx$30$^{\circ}$; \citealt{Schneider2006,Marino2016}). Hereafter, we refer to \etaTel\ A simply as \etaTel. 

Infrared (IR) data of \etaTel\ reveal an edge-on debris disk ($i\approx0^{\circ}$) with two dust temperature components of $\sim$350 K and $\sim$115 K \citep{Backman1993,Mannings1998,Chen2006,Smith2009,Chai2024}, indicating the presence of two belts of planetesimals near 22 and 26 au \citep{Smith2009}. The disk appears to be roughly coplanar within $\sim$10$^{\circ}$ with the brown dwarf companion, and the brown dwarf may gravitationally perturb planetesimals within the disk during its 71 au pericenter passage \citep{Chai2024}. 

There have been few detections of gas from the \etaTel\ disk in the literature. Herschel detected \CII\ gas but no \OI\ or H$_2$O gas \citep{Dent2013,RiviereMarichalar2014}. CO was not detected by APEX or ALMA \citep{Dent2013,Youngblood2021}.

\citet[hereafter \citetalias{Youngblood2021}]{Youngblood2021} reported a radiatively driven outflow from \etaTel's debris disk based on high-resolution ultraviolet (UV) spectroscopy revealing absorption lines of \CII, \OI, \MgII, \AlII, \SiII, \MnII, and \FeII\ near -23 \kms\ (i.e., blueshifted $\sim$20 \kms\ in the stellar rest frame). The presence of \OI* and \OI** absorption near this velocity was a major factor in the classification of this component as circumstellar, because those transitions cannot be excited in the local interstellar medium. 

However, \cite{Iglesias2023} presented compelling evidence that the absorption lines interpreted by \citetalias{Youngblood2021} as circumstellar in origin are likely actually interstellar. They presented high-resolution \ion{Ca}{2} spectroscopy of \etaTel\ and three other stars within $<$3.5\arcsec\ on the plane of the sky, including the comoving F6V star HD 181327, showing that all four sightlines exhibit \ion{Ca}{2} absorption near -23 \kms. With this evidence in mind, we revisit the interpretation of the HST/STIS spectra of \etaTel.

\section{Revisiting the evidence for a disk wind} \label{sec:ObservationsReductions}

\subsection{Summary of the UV spectra and absorption line fitting}
We briefly describe the UV spectra of \etaTel\ here and refer the reader to \citetalias{Youngblood2021} for a detailed description. \etaTel\ was observed with the Hubble Space Telescope's STIS spectrograph in two visits separated by three days (2016-Oct-03 and 2016-Oct-06) with the E140H and E230H echelle gratings. Multiple grating settings were used to provide broad wavelength coverage from 1164--1951~\AA\ and 2587--2851~\AA\ at high spectral resolving power ($\Delta \lambda = \lambda / 228\,000$~\AA~pixel$^{-1}$). No significant variability was observed between visits, with the exception of an unidentified ``mystery" absorption feature at 1594 \AA. The spectra were combined via a weighted average into a master spectrum.

The UV spectrum shows many narrow absorption lines arising from gas along the line of sight to the central star, including \CI, \CII, \NI, \OI, \MgII, \AlII, \SiII, \SII, \MnII, and \FeII. Using a Markov Chain Monte Carlo (MCMC) method, \citetalias{Youngblood2021} fit each species' absorption line to recover the radial velocity (heliocentric frame), column density, and Doppler broadening parameter for each species (see their Table 1). Many species have multiple transitions present in the spectra (e.g., \OI\ transitions at 1302.2, 1304.9, 1306.0, and 1355.6 \AA). These were fit simultaneously assuming a single radial velocity and Doppler broadening parameter for all energy levels and a single column density for each energy level. 

\subsection{Summary of the multiple velocity components detected}
Multiple velocity components were recovered for several species. Three velocity components were identified for \OI, at approximately -25 \kms, -18 \kms, and -1 \kms. Two components were detected each for \CII, \SiII, \SII, \MnII, and \FeII, at approximately -23 \kms\ and -18 \kms. For \MgII, two components were detected near -23 \kms\ and -10 \kms. Upper limits were placed on the column densities for \NI\ and \SI, and one absorption component was very tentatively detected for \CI\ (-24 \kms). 

\citetalias{Youngblood2021} summarized the above absorption line detections as a $\sim$ -23 \kms\ component and a $\sim$ -18 \kms\ component. Some scatter was observed between the different species' radial velocities, and this was attributed to uncertainties in the STIS wavelength calibration.  \MgII\ and \OI\ diverged from this framework in that \MgII\ showed absorption near -23 \kms\ and -10 \kms, and \OI\ showed absorption near -25 \kms, -18 \kms, and -1 \kms. The -1 \kms\ component was not seen in any other species.

\subsection{Reclassifying the -23 \kms\ component as interstellar}
\citetalias{Youngblood2021} interpreted the -23 \kms\ component as circumstellar and the -18 \kms\ and -10 
\kms\ components as interstellar. The \OI\ component at -1 \kms\ was labeled as unknown in origin. Although the -23 \kms\ exhibited absorption from ground energy level transitions of many species present in the interstellar medium (ISM), it was labeled as circumstellar rather than interstellar because of the presence of excited transitions of \CII\ and \OI\ (\CII*, \OI*, and \OI**) near -23 \kms. 

{ Populating O I* and O I** relies on relatively weak collisions with neutral hydrogen, whose density in the local ISM ($n$(H) $\sim$ 0.1 cm$^{-3}$; \citealt{Linsky2022a}) is well below the critical densities of the two transitions ($n_{crit}=4.7\times10^5$; \citealt{Tielens1985}).  As a result, O I* and O I** absorption is rarely if ever observed in the local ISM. On the other hand, C II* is populated with stronger collisions with electrons, whose density ($n$($e^{-}) \sim 0.1$ cm$^{-3}$; \citealt{Redfield2008_ne}) is closer to the critical density of the transition ($n_{crit}=44$; \citealt{Goldsmith2012}). Thus, populating C II* in a warm, partially ionized medium is marginally efficient. Variation in the detectability of C II* in the local ISM is attributed to spatial variations in thermal pressure and ionization \citep{Redfield2008_ne}.} 

The ground and excited levels of \OI\ triplet are occasionally seen from the Earth's atmosphere \citep{Redfield2004a}. We find that the barycentric velocity towards the \etaTel\ sightline on the dates of the STIS observations was -25.2 \kms. This value is in excellent agreement with the \OI\ -25.3 \kms\ component reported as circumstellar in \citetalias{Youngblood2021}. In that work, the slight deviation of this value from the -23 \kms\ framework was attributed to wavelength calibration uncertainty. There is probably blended terrestrial (-25 km/s) and interstellar (-23 km/s) absorption in the O I ground state transition. 
Thus, we conclude that the -23 \kms\ component is interstellar and does not trace a disk wind.

Note that we have calculated the barycentric velocity during the 49 Cet debris disk observation that yielded \OI* and \OI** detections \citep{Roberge2014} and confirm that those features are not terrestrial in origin.

\section{Is there gas in the \etaTel\ debris disk?}

With the -23 \kms\ velocity component reclassified as interstellar, we revisit the other three velocity components (-18, -10, and -1 \kms) detected toward \etaTel\ to determine whether they could be interstellar or circumstellar.

\subsection{The -18 and -10 \kms\ components: interstellar in origin}
The Ca II spectroscopy presented by \cite{Iglesias2023} revealed a $\sim$ -23 \kms\ absorption component toward \etaTel\ and three other stars nearby in projection on the plane of the sky. \etaTel\ also showed a very weak -18 \kms\ component, but its comoving star HD 181327 and the other two stars did not exhibit this component. None of the stars, including \etaTel, had a -10 or -1 \kms\ component in their Ca II spectra.

In the UV spectrum of \etaTel, the -18 \kms\ component is detected in ground-state transitions of C II, O I, Al II, Si II, and Fe II that are commonly observed in the ISM. It was not detected in the Mg II and Mn II transitions and was weakly detected in S II. The column densities of the detected species are all at least 0.5 dex lower than the those of the -23 \kms\ component \citepalias{Youngblood2021}, offering an explanation for its absence in Mn II and S II.  { S II is inherently less sensitive to lower column densities due to its lower oscillator strengths compared to the other transitions analyzed in this work.} Mn II has strong transitions in the UV but suffers from inherently lower cosmic abundance. 
The Ca II H\&K lines are also strong transitions, but Ca$^{+}$ is not the preferred ionization state of Ca in the ISM, implying lower abundance, and possibly explaining why it was possibly not detected toward HD 181327. Considering the above points, we tentatively confirm the interstellar classification of the -18 \kms\ component. High-resolution UV spectra of the stars analyzed by \cite{Iglesias2023} could resolve lingering questions about the interstellar medium in this area of the sky.

\begin{figure*}
    \centering
    \includegraphics[width=\linewidth]{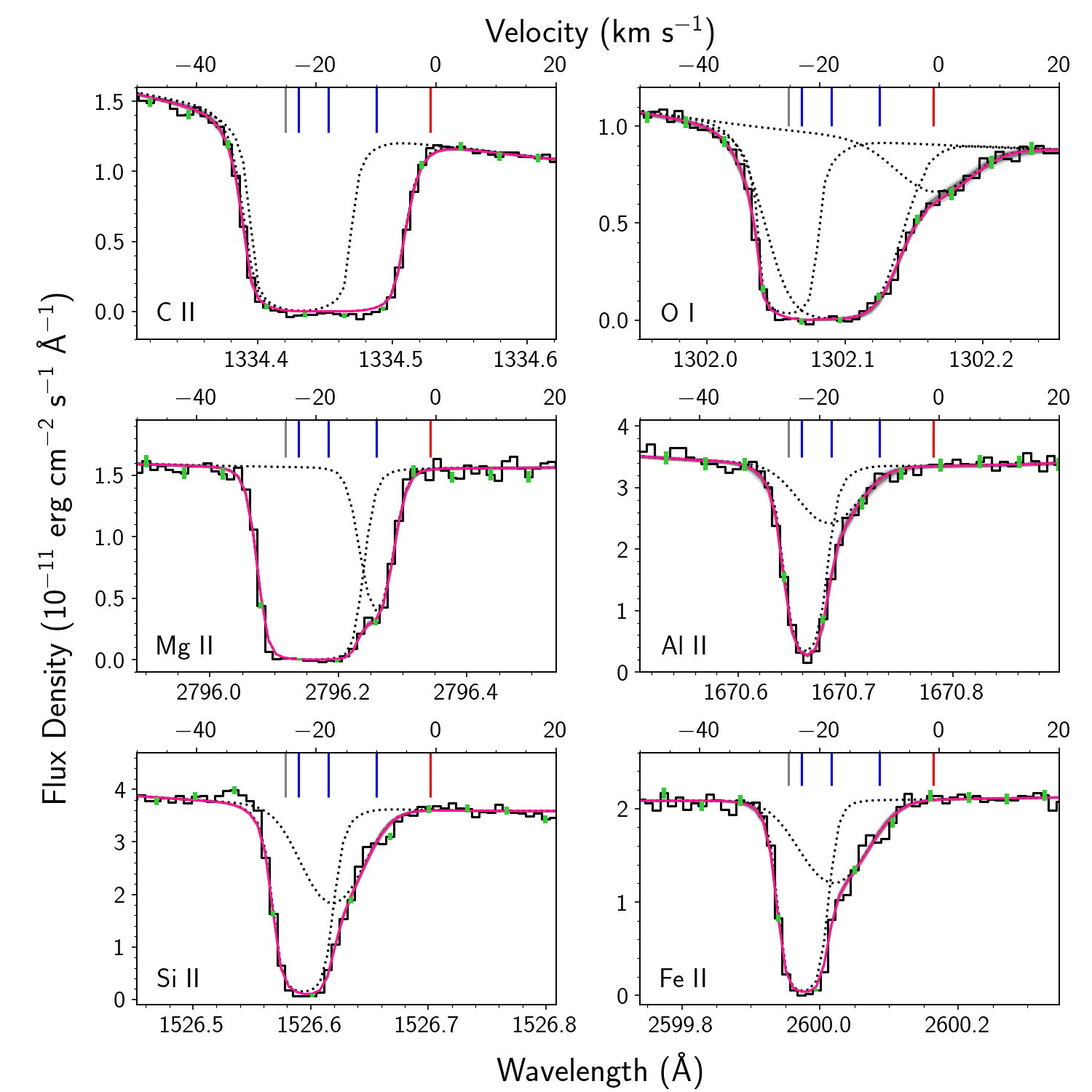}
    \caption{The STIS spectrum of \etaTel\ is shown near six strong atomic transitions where both the -23 \kms\ and -18 \kms\ (or -10 \kms\ in the case of \MgII) absorption components were clearly detected. The STIS data are shown as a black histogram with green error bars on every fifth data point for clarity. Best-fit models (pink lines with gray shading representing the uncertainty) from \citetalias{Youngblood2021} are shown along with the individual velocity components (dotted black lines). Each subplot shows the -50 \kms\ to +20 \kms\ velocity range around each transition's rest wavelength from \cite{Morton2003}. The vertical tick marks show the five velocity components discussed in this work: -25.2 \kms\ (terrestrial), -23 \kms\ (interstellar), -18 \kms\ (interstellar), -10 \kms\ (interstellar), and -1 \kms\ (possibly circumstellar).  }
    \label{fig:velocity_components}
\end{figure*}

The -10 \kms\ velocity component was only reported in \MgII\ by \citetalias{Youngblood2021}. It was deemed interstellar given its close match with the local Aql cloud's predicted radial velocity of -10$\pm$1 \kms \citep{Redfield2008}. A closer visual inspection of the UV spectrum at common ISM transitions shows evidence for an absorption component near -10 \kms\ distinct from the -18 \kms\ component for \OI, \SiII, and \FeII\ (Figure~\ref{fig:velocity_components}). \CII\ is heavily saturated, and there could be a -10 \kms\ velocity component blended with the -18 \kms\ component. Generally, the Doppler broadening parameters of the -18 \kms\ component reported by \citetalias{Youngblood2021} are greater than for the -23 \kms\ component. It is likely that the larger Doppler widths compensated for a weak -10 \kms\ velocity component not included in the fits. Thus, we conclude that the -10 \kms\ component is interstellar.

\subsection{The -1 \kms\ component: inconsistent with an interstellar origin}

The -1 \kms\ component  was only detected in \OI, and \citetalias{Youngblood2021} reported the following properties: v = $-0.8^{+4.9}_{-2.3}$ \kms, log N(\OI) = 13.5$\pm$0.2, $b$ = 8.2$^{+3.5}_{-2.1}$ \kms. { We estimate the significance of this detection as 10-$\sigma$ based on the equivalent width.} Note that there is a degeneracy between the  Doppler broadening parameter and the number of components, meaning there could be multiple narrower components. \citetalias{Youngblood2021} assumed a single component { for simplicity} and labeled the origin of this absorption as ``unknown". 

We assess the likelihood that this component originates from the ISM by comparing the column density to other values reported for the local ISM, assessing the presence or absence of other common ISM tracers near -1 \kms, and considering whether -1 \kms\ is consistent with a kinematic model of the local ISM. 

The column density for the -1 \kms\ component is on the lower end of, but still fully consistent with, the \OI\ columns reported by \cite{Redfield2004a}, such as the third components of $\kappa^1$ Cet, HR 1099, and $\alpha$ Gru. However, the third components of these three sightlines have detectable absorption from other common ISM transitions like \CII\ and \MgII, while \etaTel\ does not. In a bulk analysis of all their studied sightlines, \cite{Redfield2004a} find that \OI\ and \CII\ have similar column densities, within $\sim$0.5 dex. Additionally, the \cite{Redfield2008} kinematic model of the local ISM does not predict the absorbers near -1 \kms.  
We conclude that the -1 \kms\ component is unlikely to be interstellar. 

\subsection{Local ISM kinematic model predictions toward the \etaTel\ sightline}

We compare the \etaTel\ absorption components' radial velocities to the predictions of a kinematic model\footnote{http://lism.wesleyan.edu/LISMdynamics.html} of the local ISM \citep{Redfield2008}. The \etaTel\ sight line does not traverse any of the clouds included in the model, but passes near ($<$20$^{\circ}$) the LIC, G, Aql, and Vel clouds, which have radial velocities of $\sim$ -17, -19, -10, and -28 \kms, respectively. The reported radial velocity uncertainties are 1-1.5 \kms.

The kinematic model is constructed from sparse sight line data, so it appears these clouds have a greater spatial extent than previously thought. The LIC and G clouds could be blended into the -18 km/s component and the Aql cloud could be the -10 km/s component. The Vel cloud could be the -23 km/s component, although the 5 \kms\ discrepancy is significant. There is no known ISM cloud near the \etaTel\ sight line that has a radial velocity near -1 \kms. The Blue and Hyades cloud have predicted velocities of -6 \kms\ and -5 \kms\ but are $>$20$^{\circ}$ from the \etaTel\ sight line.

\subsection{Could the -1 \kms\ component be circumstellar?} \label{sec:circumstellar}

Stable, bound circumstellar gas in Keplerian orbit would have a radial velocity equal to the stellar radial velocity. Infalling gas (e.g., exocomets) would be redshifted, and outflowing gas would be blueshifted. \citetalias{Youngblood2021} reported a stellar radial velocity of -5.6$\pm$2.8 \kms, which is in fairly good agreement with values reported by \citep{Rebollido:2018,Rebollido:2020}, although \cite{Iglesias2023} found -0.6$\pm$1.6 \kms. Measuring the radial velocity of a young, rotationally-broadened A-type star like \etaTel\ is notoriously challenging, but it seems likely that \etaTel's radial velocity is between -5 and 0 \kms.

The -1 \kms\ component is roughly consistent with the stellar radial velocity and could be circumstellar gas on a Keplerian orbit or infalling star-grazing comets. Instead of a single broad component, there could be multiple narrow absorption components out to $\sim$ +10 \kms. Assuming exocomets would be in free-fall toward the star, we can calculate their original orbital distance \citep{Miles2016}. +5 \kms\ in the star's reference frame would correspond to $\sim$150 au and +10 \kms\ to $\sim$40 au. \citetalias{Youngblood2021} did not report any variation in \OI\ between the two HST/STIS visits that were separated by 3 days, however, the expected time baseline of exocomet variability is not well known.

To assess the likelihood that the -1 \kms\ component traces gas from colliding or evaporating planetesimals, we derive an upper limit on the C/O ratio of the gas. 
Using the same MCMC method as \citetalias{Youngblood2021}, we determine an upper limit on the C II ground state column density for the -1 \kms\ component. We fit only the ground-state line (1334.5323 \AA), permitting the free parameters to vary as follows: -5 and +5 \kms\ for the radial velocity, 10$^0$-10$^{16}$ cm$^{-2}$ for the column density, and 1--7 \kms\ for the Doppler b value. We included absorption from the -23 \kms\ and -18 \kms\ interstellar components in the model fit, but those parameters were kept fixed to the best values reported in \citetalias{Youngblood2021}. We find log N(C II) $<9.1$ (1-$\sigma$) or $<{11.4}$ (3-$\sigma$).

To infer the total carbon and oxygen columns from C II and O I, we rely on ionization balance estimates.
\cite{Fernandez:2006} presented ionization balance calculations of debris disk midplanes near 100 au. We adopt their C$^+$/C = 0.99 value for an A0V star,  although the ratio could be an order of magnitude lower in more distant orbits and/or at higher densities.  We assume O$^0$/O = 1 based on oxygen's high ionization potential energy. We derive log C/O $<$-4.4 (1-$\sigma$) or $<$-2.1 (3-$\sigma$). 

The log C/O upper limit is consistent Earth abundance ($\sim$ -2; \citealt{Allegre2001}) and solar system comets (-3 to -1.5; \citealt{Seligman2022} and references therein), but is inconsistent with solar abundance (-0.3$\pm$0.06 dex; \citealt{Lodders2003}), polluted white dwarfs ($<$0.8; \citealt{Wilson2016}; -0.5$\pm$0.23; \citealt{Sahu2025}), and the circumstellar gas of $\beta$ Pic (0.95; \citealt{Roberge2006}) and 49 Cet (0.65; \citealt{Roberge2014}).

If the -1 \kms\ component is circumstellar, the low C/O ratio compared to $\beta$ Pic and 49 Cet could arise via preferential carbon loss from the system. If CO is the parent molecule of carbon and oxygen gas in the system, we would expect log C/O = 0. \citetalias{Youngblood2021} showed that $\beta$, { the ratio of radiation pressure to gravitational force, is unity for \OI\ atoms} in this system, whereas the values for \CII\ and \CI\ are 19 and 52, respectively. \cite{Lehmets2025} report $\beta$=1.8$\pm$0.1 for \OI, 12.0$\pm$0.7 for \CII, and 53.6$\pm$3.9 for \CI, for a $T$=10\,000 K star (A0V). Thus, carbon atoms experience sufficient stellar radiation pressure to overcome the inward pull of gravity and escape the system. $\beta >$0.5 is needed to drive an outflow, so \OI\ could also be outflowing according its $\beta$ value. 
A different parent molecule such as H$_2$O could also explain the low C/O ratio. 

It is possible that the -1 \kms\ component traces carbon-poor circumstellar gas or water-rich star-grazing comets.

\subsection{Reconciling UV absorption and far-IR emission from \etaTel}

Complicating the picture of a carbon-poor disk is the Herschel PACS detection of [\CII] 158 $\mu$m emission and the absence of [\OI] 63 $\mu$m emission. \cite{RiviereMarichalar2014} report an upper limit on [\OI] emission ($F_{int}<6.2\times10^{-18}$ W m$^{-2}$) and a $\sim$3-$\sigma$ detection of [\CII] emission ($F_{int}$=2.3$\times$10$^{-18}$ W m$^{-2}$), which appears blueshifted $\sim$100 \kms\ from the stellar rest frame. The significant Doppler shift of the [\CII] emission could suggest a background object, but a visual examination of the available Herschel images of \etaTel\ shows no obvious background sources. Both \etaTel\ and its brown dwarf companion lie within a single PACS 9.4\arcsec$\times$9.4\arcsec\ spaxel, but brown dwarfs are not likely to be sources of detectable [\CII] emission and the expected radial velocity of \etaTel\ B is far less than 100 \kms\ \citep{Nogueira2024}. In addition to the brown dwarf companion, \cite{Chai2024} detected two other objects within 3\arcsec\ of \etaTel\ with JWST MIRI at 11.55 $\mu$m; both are assumed to be background galaxies although only one of them is previously known (2CXO J192251.5-542530).

The inconsistency between the UV detection of \OI\ and the far-IR non-detection of [\OI] could be attributed to gas volume densities below the critical density necessary to populate the [\OI] 63 $\mu$m line ($>$10$^5$ cm$^{-3}$; \citealt{Hollenbach1989}) leading to weak, sub-thermal emission. Indeed, Herschel detected only two debris disks in [\OI] line emission \citep{Dent2013,Brandeker2016}. In contrast, the critical density threshold for [\CII] 158 $\mu$m ($<$10 cm$^{-3}$; \citealt{Hollenbach1989}) is much more likely to be met in debris disks. The UV non-detection of \CII\ and the far-IR detection of [\CII] may indicate that the carbon gas is misaligned from the line of sight or variable in time. We confirmed that no \CII\ absorption is present in the STIS spectrum near -100 \kms.

We performed an additional check for consistency between the UV and far-IR data by comparing their derived ground-state column densities. Assuming the IR emission is optically thin and the gas is in local thermodynamic equilibrium, the column density of the ground-state is

\begin{equation}
    N = \frac{4 \pi g_l F_{int}}{A_{ul} h \nu g_u \Omega} e^{h\nu / kT_{ex}},
\end{equation}

\noindent where $\nu$ is the frequency of the transition, $A_{ul}$ is the spontaneous transition probability from the upper to lower state, $g_l$ and $g_u$ are the degeneracies associated with the lower and upper states, and $F_{int}$ is the observed integrated flux of the transition. $T_{ex}$ is the excitation temperature of the gas, and $\Omega$ is the solid angle of the emitting region.

$T_{ex}$ and $\Omega$ are unknown, so we must make broad assumptions. We assume $T_{ex} <$ 5000 K, and a range of projected disk sizes 10$^{-12}$ $<$ $\Omega$ [sr] $<$ 10$^{-10}$. $\Omega$ = 10$^{-12}$ sr corresponds to an edge-on narrow ring with inner radius 20 au, outer radius 30 au, and scale height 3 au (10\% of outer radius). $\Omega$ = 10$^{-10}$ sr corresponds to an edge-on 200 au disk with no gaps and a scale height of 20 au (10\% of outer radius). 

Under the full range of $T_{ex}$ and $\Omega$ assumptions, we find log N$_{IR}$(\CII) $>$ 16 and log N$_{IR}$(\OI) $>$ 14.3, which are inconsistent with the UV-based values log N$_{UV}$(\CII) $<$ 11.4 (3-$\sigma$) and log N$_{UV}$(\OI)=13.5$\pm$0.2. Bringing the IR-derived ground-state columns into agreement with the UV-derived ground-state columns would require large solid angles that are inconsistent with the IR-based picture of two rings of dust near 25 au \citep{Smith2009}: $>$5$\times$10$^{-10}$ sr for the \OI\ disk and $>$4$\times$10$^{-6}$ sr for the \CII\ disk. The latter value is almost three orders of magnitude larger than the PACS beam size and would represent an unphysically large disk.

\section{Summary} \label{sec:summary}
We have reviewed the evidence for and against a disk wind from the edge-on debris disk surrounding \etaTel\ based on high-resolution spectroscopy \citepalias[][\citet{Iglesias2023}]{Youngblood2021}. We confirm the interpretation of \cite{Iglesias2023} that the -23 \kms\ absorption component in the system is likely interstellar. 

With this in mind, we reconsider the -1 \kms\ absorption component detected only in \OI\ and approximately at rest within the stellar rest frame for which \citetalias{Youngblood2021} could not explain. This absorption is inconsistent with an interstellar origin and could be bound or infalling circumstellar \OI\ gas. The uncertainty surrounding the exact stellar radial velocity ($\sim$ -5 to 0 \kms) and the broad \OI\ absorption centered on -1 \kms\ prevent an exact kinematic determination. 

We derive an upper limit on the C/O ratio of the -1 \kms\ component, finding that it is consistent with Earth and solar system comets but inconsistent with the carbon-rich debris disks surrounding $\beta$~Pic and 49 Cet. The A0V central star exerts strong radiation pressure on carbon atoms and ions and weak pressure on oxygen, preferentially expelling carbon. 

Finally, we discuss whether the low C/O ratio of the -1 \kms\ component is at odds with the Herschel 3-$\sigma$ detection of [\CII] at $\sim$ -100 \kms. If the [\CII] detection is real, it may indicate that the carbon gas is misaligned with the line of sight or variable in time, because the STIS spectrum shows no \CII\ absorption near -100 \kms.

This work emphasizes the challenges associated with interpreting UV absorption toward debris disks as circumstellar because of complex ISM structure, uncertain stellar radial velocity due to rotational broadening, and terrestrial contamination.

\begin{acknowledgments}

We thank the referee for a helpful report that improved the clarity of the paper. This research is based on observations made with the NASA/ESA \emph{Hubble Space Telescope} obtained from the Mikulski Archive for Space Telescopes (MAST) at the Space Telescope Science Institute, which is operated by the Association of Universities for Research in Astronomy, Inc., under NASA contract NAS 5–26555. These observations are associated with program HST-GO-14207 (PI: A. Roberge). The specific observations analyzed can be accessed via \dataset[10.17909/t9-8q9d-6c22]{https://doi.org/10.17909/t9-8q9d-6c22}. S.P. acknowledges support from FONDECYT 1231663, ANID--Millennium Science Initiative Program NCN2024\_001 and ANID FIUF137139-USACH.

\end{acknowledgments}

\facilities{HST}
\software{astropy \citep{Robitaille2013}, IPython \citep{Perez2007}, Matplotlib \citep{Hunter2007}, NumPy and SciPy \citep{VanderWalt2011}, lyapy \citep{Youngblood2016}, emcee \citep{Foreman-Mackey:2013}.}

\bibliography{main_arxiv.bbl}{}
\bibliographystyle{aasjournalv7}

\end{document}